\documentclass[reprint,aps,prd,floats,floatfix,amsmath,amssymb,amsfonts,nofootinbib,longbibliography,superscriptaddress]{revtex4-2}

\usepackage{graphicx}
\usepackage{dcolumn}
\usepackage{bm}
\usepackage{natbib}
\usepackage[export]{adjustbox}

\usepackage[hyperindex,colorlinks]{hyperref}

\renewcommand{\vec}{\boldsymbol}

\newcommand*{\mathabxbfamily}{\fontencoding{U}\fontfamily{mathb}\selectfont}
\DeclareFontFamily{U}{mathb}{\hyphenchar\font45}
\DeclareFontShape{U}{mathb}{m}{n}{
      <5> <6> <7> <8> <9> <10> gen * mathb
      <10.95> mathb10 <12> <14.4> <17.28> <20.74> <24.88> mathb12
      }{}
\newcommand*{\Earth}{{\text{\mathabxbfamily\char"43}}}
\newcommand*{\Moon}{{\text{\mathabxbfamily\char"4B}}}

\begin{document}

\title{Lagrangian Extensions of Newtonian Gravity constrained by Solar System tests}

\author{Pedro H. Dalpr\'a}
\affiliation{%
Programa de P\'os-Graduação em F\'isica (PPGF\'is),  Universidade Federal do Esp\'irito Santo (UFES)\\
Av. Fernando Ferrari, 540, CEP 29.075-910, Vit\'oria, ES, Brasil.}%
\author{J\'ulio C. Fabris}
\affiliation{%
Departamento de F\'isica,  Universidade Federal do Esp\'irito Santo (UFES)\\
Av. Fernando Ferrari, 540, CEP 29.075-910, Vit\'oria, ES, Brasil.}%
\author{Hermano Velten}%
\affiliation{%
Departamento de F\'isica, Universidade Federal de Ouro Preto (UFOP), Campus Universit\'ario Morro do Cruzeiro, 35.400-000, Ouro Preto, MG, Brasil}%
\affiliation{%
Instituto de Astronomía Teórica y Experimental, CONICET-UNC, X5000BGR, Córdoba, Argentina}%
\author{Júnior D. Toniato}%
\email[E-mail:~]{junior.toniato@ufes.br}
\affiliation{%
Departamento de Qu\'imica e F\'isica,
Universidade Federal do Esp\'irito Santo (UFES) -- Campus Alegre.\\
Alto Universitário, s/nº, 29.500-000, Alegre, ES, Brasil}%
\collaboration{%
N\'ucleo de Astrofísica e Cosmologia \ \includegraphics[scale=0.07, valign=c]{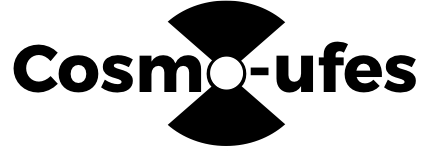}\,
Universidade Federal do Esp\'irito Santo
}%
\homepage[Visit:~]{www.cosmo-ufes.org}
\noaffiliation

\date{\today}

\begin{abstract}
We explore an extension to Newtonian gravity through a generalised Lagrangian function with the introduction of a second dynamical scalar field. Building on previous research into gravity with variable gravitational coupling, the work derives the complete field equations and applies a weak-field approximation. This leads to an effective post-Newtonian gravitational potential that includes key aspects of relativistic theories. The resulting N-body equations of motion highlight differences among inertial and gravitational masses, which can constrain the theory's free parameter through data from the Nordtvedt effect. By employing the method of osculating orbits for a two-body system, the study calculates the secular variation of the orbital pericenter and aligns this with the latest data on Mercury's perihelion shift, for another observational constraint on the model. Furthermore, a few examples of theories are discussed.
\end{abstract}

\maketitle

\section{Introduction}

The rise of Newtonian mechanics represented a unique revolution for physics and science in general. For nearly three centuries, so-called classical physics, based on the theoretical framework introduced by Newton, has been the basis for countless scientific and technological advances. In the field of gravitation, this impact was no different: the theory of universal gravitation, presented in his \textit{Principia}, constituted a model capable of satisfactorily explaining phenomena within the Solar System realm and provided results that paved the way for discoveries in celestial mechanics. A historical and remarkable example is the application of perturbative techniques in Newtonian gravity to the hypothesis of a supposed perturber to Uranus' orbit, leading to the discovery of Neptune \cite{smith1989cambridge}.

Around the mid-19th century, astronomers observed that Mercury's perihelion - the point in its orbit where the planet is closest to the Sun - was progressively shifting over time, advancing at a rate of $574.1$ seconds of arc per century \cite{poisson_will_2014}. According to the two-body dynamics described by Kepler, the perihelion should remain fixed. However, it seemed clear that the shifting was caused by the gravitational influence of the other planets, which would slightly perturb Mercury's orbit.

Le Verrier calculated the contribution of each planet to the advance of perihelion, but the results showed a discrepancy of $42.9$ seconds of arc per century, which could not be explained by known interactions \cite{poisson_will_2014}. Only in 1915, with Einstein's theory of general relativity, did a formulation emerge that was consistent with observational data.

The theory of general relativity has established itself as the dominant framework for describing gravitational interactions. Despite its success in accounting for a wide range of experiments at the solar-system scale, cosmological scenarios have challenged its domain of validity. Consequently, alternative theories have been developed, motivated by the intriguing evidence of a dark sector in the universe—comprising dark matter and dark energy—which together account for $ \sim 95\%$ of the cosmic energy budget \cite{Wands:1993uu,Capozziello:2011et,Clifton:2011jh,Joyce:2016vqv,Koyama:2015vza,Nojiri:2010wj,Nojiri:2017ncd}. However, many relativistic gravitational theories may have classical limits different from those of a Newtonian theory. Hence, screening mechanisms are invoked to restore the locally well-tested general relativity results \cite{Kimura:2011dc, Hinterbichler:2010es}. See also Ref. \cite{March:2021mqu} for an interesting application related to the present work. Moreover, a common feature of relativistic/metric theories of gravity is the variation of the gravitational coupling. Concerning the latter, it is worth mentioning that the first proposals to incorporate non-relativistic varying $G$ theories were presented in Refs. \cite{Landsberg:1975, McVittie:1978, Duval:1990hj}. For example, Ref. \cite{McVittie:1978} introduced the idea that the constant $G$
 can be replaced by a time-dependent gravitational coupling, $G(t)$, in the Poisson equation, with no dynamical equation governing this new function. Then, its temporal behaviour must be imposed ad hoc. For example, a simple choice in this context is to adopt Dirac’s proposal $G \propto t^{-1}$ \cite{Dirac:1937ti,Dirac:1938mt}.

While general relativity and its various extensions stand as the standard way of understanding gravitational phenomena, one can still effectively analyse many astrophysical and cosmological systems via Newtonian gravity. This approach offers huge mathematical and numerical simplicity that can be quite appealing in comparison to general relativity techniques. In this context, extending Newtonian gravity by incorporating novel features that reflect the intricate relativistic effects observed in modern gravitational theories can be understood as a significant contribution. Such an upgrade would prove invaluable for the exploration of weak-field systems. It is also desirable to design a non-relativistic gravitational theory with its own mechanism to generate the dynamical evolution of any extra degree of freedom, e.g., a new scalar field, rather than relying on {\it ad hoc} assumptions for that. Covariant theories known as scalar-tensor theories provide such a mechanism where gravity is mediated both by the metric field and a new scalar field, with the Brans-Dicke theory the prototype of this class \cite{Brans:1961sx} and the so-called Horndeski family of theories representing the most general framework for them \cite{Horndeski:1974wa,Kobayashi_2019}. 

Motivated by these ideas, Ref. \cite{jjtv} proposes a new Newtonian theory that incorporates a varying gravitational coupling and discusses its implications for cosmology. Subsequently, the authors examined stellar stability \cite{Fabris:2021qkp} and celestial mechanics \cite{Escorcio:2023}, both of which provided observational constraints on the theory's free parameter. In summary, these references proposed to answer the following issue: is there a Newtonian-like version of relativistic scalar-tensor theories?

Inspired by the works mentioned above, the present study represents an effort to investigate more general extensions of the Newtonian Lagrangian for the gravitational field that remain consistent with observational data.
The general theoretical setup is reviewed in the next section. We then proceed to calculate the corrections to the Newtonian gravitational potential within a weak-field approximation in the subsequent section. The equations of motion are presented in section \ref{sec:eq-motion} and the equivalence principle discussed in section \ref{sec:ep}. Applications to the pericenter advance are performed in section \ref{variat-elem}.  We conclude in the final section.

\section{The standard and the extended Newtonian gravity}

The Poisson equation of the Newtonian theory of gravity can be obtained from the Lagrangian:
\begin{equation}
\mathcal L = \frac{\vert\vec\nabla U\vert^2}{8\pi G} - \rho U,
\end{equation}
where $\rho$ is the density of matter, $U$ is the Newtonian gravity potential and $G$ is the constant gravitational coupling. Inserting this Lagrangian into the Euler-Lagrange equation,
\begin{equation}
\vec\nabla\cdot\frac{\partial\mathcal L}{\partial\vec\nabla U} - \frac{\partial\mathcal L}{\partial U} = 0,
\end{equation}
the Poisson equation is directly obtained,\footnote{Rigorously, $U$ is defined as the negative of the ordinary Newtonian potential. The reason why we are using this convention is to be consistent with traditional post-Newtonian notation.}
\begin{equation}
\nabla^2 U = -4\pi G\rho.
\end{equation}
The well-known solution of the  equation above, obtained by Green method, is
\begin{equation}\label{newt}
    U=G\int\frac{\rho(x')}{\vert \vec{x}-\vec{x}{}'\vert}\,d^3x'.
\end{equation}

The proposed extension of Newtonian gravity under investigation is formulated with two scalar fields, designated as $\psi$ and $\sigma$. This extension is predicated upon the following Lagrangian,
\begin{align}\label{LagrangianG}
\mathcal{L} =& \, \frac{k(\sigma)}{8\pi G_0}\,\vert\vec{\nabla}\psi\vert^2 \notag\\[1ex]
&-\frac{\omega}{8\pi G_0}\left[f(\psi,\sigma)\dot{\sigma}^2
-g(\psi,\sigma)\vert\vec{\nabla}\sigma\vert^2\right] + \rho h(\sigma)\psi,
\end{align}
where $\omega$ is a dimensionless constant and $G_0$ is a constant with the same units as $G$.  The over-dot indicates the total time derivative. The terms $k(\sigma)$, $f(\psi, \sigma)$, $g(\psi, \sigma)$ and $h(\sigma)$ play the role of coupling functions. At the Lagrangian level, the connection with the standard Newtonian theory can be restored by simultaneously setting $\sigma$ as a constant and with the identification $G_0=Gh/k$, which gives $\psi=-U$. Therefore, one can interpret $\psi$ as an equivalent of the ordinary Newtonian potential, while $\sigma$ is a new dimensionless dynamical field.

The field equations of the theory are obtained by applying the Lagrangian to the Euler-Lagrange equations,
\begin{align}
\label{EulerLagrange1}
\vec\nabla\cdot\frac{\partial\mathcal L}{\partial\vec\nabla\psi} &- \frac{\partial\mathcal L}{\partial\psi} = 0,\\[1ex]
\label{EulerLagrange2}
\vec\nabla\cdot\frac{\partial\mathcal L}{\partial\vec\nabla\sigma} &+ \frac{d}{dt}\frac{\partial\mathcal L}{\partial\dot\sigma} - \frac{\partial\mathcal L}{\partial\sigma} = 0.
\end{align}
In this way, the following equations are obtained:
\begin{align}
k(\sigma)\nabla^2\psi &+ k_\sigma\vec\nabla\sigma\cdot\vec\nabla\psi \notag \\[1ex]
&+ \frac{\omega}{2}\left(f_{\psi}\dot{\sigma}^2 - g_{\psi}\vert\vec{\nabla}\sigma\vert^2\right) = 4 \pi G_0 h(\sigma)\rho,\label{eqpsi} \\[2ex]
g(\psi, \sigma)\nabla^2\sigma &- f(\psi, \sigma)\ddot{\sigma} - k_{\sigma} \frac{\vert\vec{\nabla}\psi\vert^2}{2\omega} + \frac{g_{\sigma}\vert\vec{\nabla}\sigma\vert^2}{2} \notag \\[1ex]
- f_\psi\dot\psi\dot\sigma &+ g_{\psi}(\vec{\nabla}\psi\cdot\vec{\nabla}\sigma) - \frac{f_{\sigma}\dot{\sigma}^2}{2} = \frac{4\pi G_0}{\omega}\rho\psi h_{\sigma},\label{eqsigma}
\end{align}
such that $k_\sigma$, $g_\sigma$, $g_\psi$, $f_\sigma$ and $h_\sigma$ indicate partial derivatives of the functions with respect to the field shown in the subscript, i.e. given $\alpha=\left\{\sigma,\psi \right\}$,
\begin{equation}
    F_\alpha\equiv\frac{\partial F}{\partial\alpha}\quad \text{and} \quad F_{\alpha\beta}\equiv\frac{\partial^2 F}{\partial\beta\partial\alpha}.
\end{equation}
Now, at the dynamical equations level, one can see that the Newtonian limit of the theory, \textit{id est}, the Poisson equation, is obtained both when $\sigma$ is a constant and $\omega\rightarrow\infty$ simultaneously.

It should be stressed that the foregoing system of equations cannot be consistently interpreted as the non-relativistic limit of any currently established, fully covariant scalar–tensor theory of gravitation. Since the Newtonian regime is recovered only through particular limiting choices of the parameters $\sigma$ and $\omega$, in analogy with the Brans–Dicke construction, one can at best assert a structural similarity to the Brans–Dicke framework. A genuine scalar–tensor theory that naturally reproduces this non-relativistic dynamics in an appropriate low-energy or weak-field limit has yet to be formulated.

\section{The weak-field approximation}
In this section, we will analyse the structure of the effective gravitational potential using a weak-field approximation. Within this framework, we expand the scalar fields in series,
\begin{align}
    \psi &=\psi_0 +\psi_1 + \psi_2 + ... \\
    \sigma &=\sigma_0 +\sigma_1 + \sigma_2 + ... \,,
\end{align}
where $\psi_2\ll\psi_1\ll\psi_0$ and $\sigma_2\ll\sigma_1\ll\sigma_0$. The background terms, $\psi_0$ and $\sigma_0$, are considered to be exact solutions of the field equations, and are treated as constants. The latter feature is consistent with the study of a local system where, within its length scale, the global background field is approximately invariant. For example, the background field can represent the gravitational influence of the galaxy's central black hole in the Solar System, whereas higher-order fields in the expansion represent the gravitational influence of the Sun and planets. Moreover, the background field can be assumed to vary over a very long time span; thus, it can be assumed that each time derivative of the perturbed fields increases the perturbative order by one. In summary, for instance: $\psi_1\sim\mathcal O(1)$ and $\dot\psi_1\sim\mathcal O(2)$.

Our goal now in the following is to solve the field equations, order by order.

\subsubsection{First order perturbations}
Applying the approximation scheme proposed above to the field equations and retaining only first-order terms, one obtains
\begin{align}
    \nabla^2\psi_1&=\frac{4\pi G_0h_0}{k_0}\,\rho,\\[1ex]
    \nabla^2\sigma_1&=\frac{4\pi G_0h_\sigma}{\omega g_0}\,\psi_0\rho,
\end{align}
where subindex $0$ in the free functions indicates they are being evaluated at the background.
The solutions of these equations are proportional to the ordinary Newtonian potential $U$,
\begin{align}
    \psi_1=-\frac{G_0h_0}{Gk_0}\,U,\quad \sigma_1 = -\frac{G_0\psi_0h_\sigma}{G\omega g_0}\,U.
\end{align}

\subsubsection{Second order}
The second-order field equations are
\begin{align}
    \nabla^2\psi_2 &= \frac{\omega g_\psi}{2k_0}(\vec\nabla\sigma_1)^2 - \frac{k_\sigma}{k_0}\vec\nabla\sigma_1\cdot\vec\nabla\psi_1 \notag \\[1ex]
    &\quad + \frac{4\pi G_0\rho\sigma_1}{k_0}\left(h_\sigma - \frac{h_0k_\sigma}{k_0}\right),\label{eqpsi2}\\[2ex]
    \nabla^2\sigma_2 &= k_\sigma\frac{(\vec\nabla\psi_1)^2}{2\omega g_0} - g_\sigma\frac{(\vec\nabla\sigma_1)^2 + 2\sigma_1\nabla^2\sigma_1}{2g_0} \notag \\[1ex]
    &\quad - \frac{g_\psi}{g_0}\Big[(\vec\nabla\psi_1\cdot\vec\nabla\sigma_1) + \psi_1\nabla^2\sigma_1\Big]\notag\\[1ex]
    &\quad + \frac{4\pi G_0\rho\psi_0h_{\sigma\sigma}\sigma_1}{\omega g_0} + \frac{4\pi G_0\rho\psi_1h_\sigma}{\omega g_0}.\label{eqsig2}
\end{align}
From the first-order solutions, we derive the post-Newtonian gravitational potential $\Phi_2$, which arises in the PPN formalism, and satisfies the following equation 
\begin{equation}
    \nabla^2\Phi_2=-4\pi G\rho U.
\end{equation}
It is again trivial to find that the integral solution for this equation can be expressed as
\begin{equation}
    \Phi_2=G\int\frac{\rho(x')U(x')}{\vert \vec{x}-\vec{x}{}'\vert}\,d^3x'.
\end{equation}
Additionally, there is a helpful relation for integrating equations \eqref{eqpsi2} and \eqref{eqsig2}:
\begin{equation}
(\vec\nabla U)^2 = \nabla^2\left(\frac{U^2}{2} - \Phi_2\right).
\end{equation}
Ultimately, one finds,
\begin{widetext}
\begin{align}
    \psi_2 &= \frac{\psi_0h_\sigma}{\omega k_0g_0}\left(\frac{G_0}{G}\right)^2 \bigg[\left(\frac{\psi_0h_\sigma}{2g_0}g_\psi - \frac{h_0k_\sigma}{k_0}\right)\frac{U^2}{2}
    + h_\sigma\left(1 - \frac{\psi_0}{2g_0}g_\psi\right)\Phi_2\bigg],\\[2ex]
    \sigma_2 &= \frac{1}{\omega g_0}\left(\frac{G_0}{G}\right)^2 \bigg[\left(\frac{h_0^2k_\sigma}{2k_0^2} - \frac{\psi_0g_\psi h_0 h_\sigma}{g_0k_0} - \frac{\psi_0^2g_\sigma h_\sigma^2}{2\omega g_0^2} \right)\frac{U^2}{2} +
     \left(\frac{h_0h_\sigma}{k_0} + \frac{\psi_0^2h_\sigma h_{\sigma\sigma}}{\omega g_0} - \frac{h_0^2k_\sigma}{2 k_0^2} - \frac{\psi_0^2h_\sigma^2g_\sigma}{2\omega g_0^2} \right)\Phi_2\bigg].
\end{align}
\end{widetext}

Finally, one is now able to formulate the approximated effective gravitational potential of the theory, namely
\begin{align}
    U_{\rm eff}&=-\psi h(\sigma)\notag \\[1ex]
    &\approx -\psi_0h_0 + U - \kappa U^2 - 2\kappa \Phi_2, \label{Ueff}
\end{align}
with
\begin{align}\label{kappa}
\kappa = \ &\frac{\psi_0 h_\sigma^2 k_0}{(\omega g_0 h_0^2 + \psi_0^2 h_\sigma^2 k_0)^2}\bigg( \omega g_0 h_0 - \frac{\psi_0 \omega g_\psi h_0}{4} \notag \\[1ex]
&\qquad  - \frac{\psi_0^2 g_\sigma h_\sigma k_0}{4 g_0} + \frac{\psi_0^2 h_{\sigma\sigma} k_0}{2} - \frac{\omega g_0 h_0^2 k_\sigma}{4k_0h_\sigma} \bigg).
\end{align}
In equation \eqref{Ueff}, it was required that the theory have a proper Newtonian limit, which is achieved by fixing $G_0$ through the condition
\begin{equation}
\frac{G_0}{G}\left(\frac{\psi_0^2h_\sigma^2}{\omega g_0} + \frac{h_0^2}{k_0}\right) = 1.\label{g0}
\end{equation}


It is important to note that both second-order potentials, \(U^2\) and \(\Phi_2\), appear in the post-Newtonian expansion of general relativity. It is well established that these potentials have a direct impact on the time variation of the orbital parameters of a binary system, such as the longitude of the periastron \cite{poisson_will_2014}. Therefore, in the next section, we will derive the equations of motion for an N-body system, which will be helpful to analyse orbital perturbations.

\section{Equation of motion of massive bodies} \label{sec:eq-motion}

We are interested in deriving the equation of motion of a certain body $A$ within an N-body system. The center-of-mass acceleration of a specific body reads
\begin{equation}\label{eqForce}
\vec{a}_A=\frac{1}{m_A}\int_A\rho(t,\vec{x})\,\vec\nabla U_{\rm eff}\,d^3x,
\end{equation}
where $m_A$ is its inertial mass.
In the above expression, the domain of integration is a fixed volume $V_A$, which is bigger than the volume occupied by body $A$ but also small enough such that it does not intersect any other body of the system. This is feasible if we assume the bodies are far apart from each other.

To compute the gravitational potential within the integral, we need to consider the contributions from all bodies in the system, including body $A$. For that reason, we will rearrange the potentials as follows,
\begin{align}
    U=U_A + U_{{\rm ext}}& &\text{and}& &\Phi_2 = \Phi_{2,A} + \Phi_{2,{\rm ext}},
\end{align}
where $U_A$ and $\Phi_{2, A}$ stand for these potentials as generated exclusively by body $A$, while $U_{{\rm ext}}$ and $\Phi_{2,{\rm ext}}$ are the contributions due to the remaining bodies of the system. Both potentials are calculated as in Eq. \eqref{newt}, with potentials from $A$ integrated over $V_A$ and external potentials summed over the volumes surrounding the other bodies. By assuming a symmetric density around the center of mass, only the external parts contribute to the acceleration. Moreover, considering the bodies are widely separated, the external potentials can be treated as independent of the integration variables as a first approximation. After the above considerations, one can write
\begin{align}
    a_A^j =  \partial_jU_{{\rm ext},A} &-2\kappa\,\partial_j\Phi_{2\,{\rm ext},A} \notag\\[1ex]
    &+2\kappa \left(2\frac{\Omega_A}{m_A} - U_{\rm ext,A}\right)\partial_jU_{\rm ext,A}\,,\label{force}
\end{align}
where $\partial_j\equiv \partial/\partial x^j$ and
with $\Omega_A$ being the gravitational energy of body $A$.

Once we assume the bodies are well separated, the external potentials can be expanded in a Taylor series (this is a standard procedure with details provided in Ref. \cite[p. 437]{poisson_will_2014}), which gives
\begin{align}
    U_{{\rm ext}} &= \sum_{B\neq A}\frac{Gm_B}{r_{AB}},\\[1ex]
    \partial_jU_{{\rm ext}} &= -\sum_{B\neq A}\frac{Gm_B}{r_{AB}^2}\,\hat{r}_{AB}^j,\\[1ex]
    \partial_j\Phi_{2,{\rm ext}} &= 2\sum_{B\neq A}\frac{G\Omega_B}{r_{AB}^2}\,\hat{r}_{AB}^j - \sum_{B\neq A}\frac{G^2m_Am_B}{r_{AB}^3}\,\hat{r}_{AB}^j \notag\\[1ex]
    & \qquad \qquad - \sum_{B\neq A}\sum_{C\neq A,B}\frac{G^2m_Bm_C}{r_{AB}^2r_{BC}}\,\hat{r}_{AB}^j.
\end{align}
Here and after, it is used the notation $\vec{r}_{AB}=\vec{r}_A - \vec{r}_B$, and $\hat{r}_{AB}=\vec{r}_{AB}/r_{AB}$. Substituting the terms into the equation \eqref{force} yields
\begin{widetext}
\begin{align}\label{acceleration}
a_A^j = -\sum_{B\neq A}\frac{G\mathcal{M}_A\mathcal{M}_B}{m_A \,r_{AB}^2}\hat r_{AB}^j + 2\kappa\sum_{B\neq A}&\frac{G^2\mathcal{M}_A\mathcal{M}_B}{m_A \,r_{AB}^3}\Big(\mathcal{M}_A + \mathcal{M}_B\Big)\hat r_{AB}^j \notag \\[1ex]
&\quad + 2\kappa\sum_{B\neq A}\sum_{C\neq A,B}\frac{G^2\mathcal{M}_A\mathcal{M}_B\mathcal{M}_C}{m_A \,r_{AB}^2}\left(\frac{1}{r_{AC}} + \frac{1}{r_{BC}}\right)\hat r_{AB}^j,
\end{align}
\end{widetext}
where we have considered the mass redefinition
\begin{align}
\mathcal M = m\left(1 + 4\kappa\frac{\Omega}{m}\right),\label{mass}
\end{align}
such that $\mathcal{M}$ is the gravitational mass of the body. Moreover, terms of order $\kappa\Omega^2/m^2$ are being neglected.

Consequently, a distinction between inertial and gravitational masses arises in this extension of Newtonian theory. We will discuss this aspect in the next section.

\section{The equivalence principle}
\label{sec:ep}

The Equivalence Principle is one of the cornerstones of General Relativity. Yet, as discussed in Ref. \cite{Damour:2012rc}, contrary to the status of principles of physics such as energy conservation or the principle of least action, it should not be classified among the fundamental ones. It emerged as a generalisation of the empirical evidence that free-falling neutral bodies experience the same acceleration in an external gravitational field. From a more refined and technical perspective, this principle can be understood as a consequence of one of the two foundational postulates of general relativity: the universal coupling between matter and gravity, realised by replacing the Minkowski metric of special relativity with a curved spacetime metric. 

 To discuss the equivalence principle in extensions of Newtonian gravity, it is essential to clearly present the various types of mass definitions that characterise gravitational interactions. Therefore, we write the Newtonian acceleration of a body $A$ within an $N$-body system as follows,
\begin{equation}\label{acceleration-N}
    \vec{a}_A = -\frac{(\mathcal{M_P})_A}{(\mathcal{M_I})_A}\sum_{B\neq A}\frac{G(\mathcal{M_A})_B}{r_{AB}^2}\vec{\hat r}_{AB}.
\end{equation}
The above expression ordinarily introduces the definition of active gravitational mass $({\cal M_A})$: the amount of mass generating a gravitational potential; {passive gravitational mass} $({\cal M_P})$: the proportionality factor relating gravitational force with the gradient of the potential; and {inertial mass} $({\cal M_I})$: the general mass term given by the ratio between force and acceleration (Newton's second law). 

In Newton's ordinary gravitational theory, it follows that $\mathcal{M_I} = \mathcal{M_P} = \mathcal{M_A}$. Theories that have inertial mass equal to passive gravitational mass guarantee the validity of the equivalence principle, i.e., regardless of their mass or internal structure, all bodies are equally accelerated by the same gravitational field. Theories with equal passive and active gravitational masses satisfy conservation of total momentum in the absence of external forces acting on the system.

In the extended Newtonian gravity treated here, inertial mass is given by the material density integral, i.e. ${\cal M_I}=m$, and
active and passive gravitational masses are equivalent, i.e.
\begin{equation}\label{gravmass}
\mathcal{M_P} = \mathcal{M_A} = \mathcal{M}.
\end{equation}
Therefore, total momentum conservation of self-gravitating systems is preserved. However, we still observe a deviation from the equivalence principle, as this gravitational mass depends on the body's gravitational energy and differs from inertial mass [cf. \eqref{mass}],
\begin{equation}
    m\neq \mathcal{M}.
\end{equation}
Once theories within this class permit violations of the principle of free fall, it becomes necessary to undertake a more detailed analysis of the constraints that experimental measurements and observational data can impose on the model.

\subsection{Test of the universality of free fall}
The equality between inertial and gravitational mass has been tested in various environments. Specifically, if test particles with different compositions fall at different rates in an external gravitational field, it would violate the weak equivalence principle (WEP), which states that inertial and gravitational masses are equal. Conversely, violations of the strong equivalence principle (SEP) occur when extended bodies, which have different gravitational self-energies, are accelerated differently in a gravitational field.

The WEP has been tested by torsion balances in the laboratory, as in the one explored in Ref. \cite{Adelberger:2003}, but also in a space experiment, the MICROSCOPE mission \cite{microscope}. Both procedures deal with small test masses with similar gravitational self-energies, which makes those tests unsuitable for us.


The Lunar Laser Ranging (LLR) data were designed to measure the Earth-Moon distance and possess significant appeal due to their high and distinctive gravitational self-energies.
An eventual violation of SEP would manifest itself through the Nordtvedt Effect, which has an observational signature of oscillations in the lunar orbit. This effect is characterised by the relative acceleration between the Earth and the Moon, due to the gravitational interaction with the Sun,
\begin{align}
\Delta &= \frac{a_\Moon - a_\Earth}{a_\Earth}\approx \left(\frac{\mathcal M_P}{\mathcal{M_I}}\right)_\Earth - \left(\frac{\mathcal M_P}{\mathcal{M_I}}\right)_\Moon,
\end{align}
where $\Earth$ and $\Moon$ represent the Earth and the Moon, respectively. The most recent estimations of $\Delta$ using LLR data can be found in Refs. \cite{Hofmann_2018} and \cite{Viswanathan2018}, with both delivering the following result
\begin{equation}
    \Delta \lesssim 10^{-14}.
\end{equation}

For the generalised Newtonian gravity explored in this work, it is shown that [cf. \eqref{mass}],
\begin{equation}
\Delta \approx 4\kappa\left[\left(\frac{\Omega}{m}\right)_\Earth - \left(\frac{\Omega}{m}\right)_\Moon\right].
\end{equation}
The values of self-gravitational energy per unit of mass in the Earth-Moon system are read as \cite{Williams1996}
\begin{align}
    \left(\frac{\Omega}{m}\right)_{\Earth}
    &\cong 41.76\times10^{6} \, \text{[m/s]²},\\[2ex]
    \left(\frac{\Omega}{m}\right)_{\Moon}
    &\cong 1.71\times10^{6} \, \text{[m/s]²},
\end{align}
In this scenario, one finds the constraint
\begin{equation}
    \kappa \lesssim 10^{-22} \ \mbox{(m/s)}^{-2}.\label{sep-constraint}
\end{equation}

\section{Variations of orbital elements of a binary system}\label{variat-elem}
Considering a system of two self-gravitating bodies, their relative acceleration, $\vec a=\vec{a}_1-\vec{a}_2$, is directly obtained from Eq. \eqref{acceleration}, yielding
\begin{equation}
\vec{a}=-\frac{Gm^*}{r^2}\,\vec{\hat r}+2\kappa\frac{G^2m^2}{r^3}\vec{\hat r},\label{acceleration-bs}
\end{equation}
with $\vec r=\vec r_1 - \vec r_2$, $r=|\vec r|$, $\vec{\hat r}=\vec r/r$, $m=m_1+m_2$ and 
\begin{equation}
    m^*\equiv m\,\frac{\mathcal{M}_1\mathcal{M}_2}{m_1 m_2}.
\end{equation}
It is noteworthy that \eqref{acceleration-bs} closely resembles the force derived from the Manev potential, originally proposed in the 1920s \cite{Maneff1925,Diacu}, which constituted an early attempt to reproduce general relativistic effects through a suitably modified Newtonian potential. It is also worth noting that a $1/r^3$ correction to the Newtonian gravitational force between two masses likewise arises within effective field theory frameworks employed to study quantum modifications of the Newtonian force law (see, for example, Refs.~\cite{Battista:2014oba,Battista:2020qqp} and references therein).

The $m^*$ is the mass term that controls the structure-dependent effects resulting from the difference between inertial and gravitational masses.
However, $m^*$ also plays the role of the Keplerian mass, i.\,e. it is the single measurable mass determined through Kepler's third law. As a result, the corrections related to internal structure are not observable in a binary system, and we can safely omit the ${}^*$ symbol and express $\vec a$ in terms of $m$ solely.


With the disturbance in the relative acceleration of a binary system, the trajectory \( r(\phi) \) will no longer be a perfect ellipse. However, it can still be treated as an instantaneous ellipse, with its semi-latus rectum \( (p) \) and eccentricity \( (e) \) changing over time.  Additionally, the angular positional parameters will also be time-dependent, namely, the pericenter argument $(\omega)$, the longitude of the ascending node $(\Omega)$, and the orbital inclination $(\iota)$. The method of osculating orbits addresses these variations and establishes a direct relationship between them and the components of the perturbed acceleration (see Ref. \cite[p.\,160]{poisson_will_2014}). Once Eq. \eqref{acceleration-bs} shows deviations from the Newtonian case in the radial direction only, it follows that just $e$ and $\omega$ undergo time variations. When described in terms of the polar angle, the changes in both parameters read
\begin{align}
    \frac{dw}{d\phi} =& \ - \frac{p^2}{eGm}\frac{\cos \phi}{(1+e\cos \phi)^2}\,{\cal R},\label{dw}\\[1ex]
    \frac{de}{d\phi} =& \  \frac{p^2}{Gm}\frac{\sin \phi}{(1+e\cos \phi)^2}\,{\cal R},\label{de}
\end{align}
where
\begin{equation}
    \mathcal R = 2\kappa\frac{G^2m^2}{r^3}.
\end{equation}

Using the relation $r = {p}/(1+e\,\cos{\phi})$, we integrate the expressions \eqref{dw}--\eqref{de} over a complete orbital period to obtain
\begin{align}
    \Delta w =& \ -\,\frac{2\kappa\pi Gm}{a(1-e^2)},\\[1ex]
    \Delta e =& \ 0.
\end{align}
The variation per orbit can be converted into a variation per time by dividing $\Delta w$ by an orbital period $P$. In this process, we can eliminate the period through Kepler's third law, i.e. $P^2=4\pi^2 a^{3}/{Gm}$, with $a=p/(1-e^2)$ the semi-major axis of the orbital ellipse.\footnote{In principle, Kepler's third law would be modified due to the new gravitational force, but since $\Delta w$ is already of post-Newtonian order, the corrections are negligible.} The result is
\begin{equation}\label{wsec}
    \left(\frac{dw}{dt} \right)_{\rm sec} = -\frac{\kappa(Gm)^{3/2}}{a^{5/2}(1-e^2)}.
\end{equation}

\subsection{Observational Constraints}\label{constraints}

The shift in the orbital pericenter of a binary system presents an opportunity to derive observational limits for the parameter $\kappa$. One of the most iconic demonstrations of the advancement of an orbital pericenter is the celebrated case of Mercury. This pivotal experiment stands as a classic in gravitational physics and played an essential role in the triumph of general relativity. The argument of Mercury's perihelion experiences a slow but notable shift, influenced by the gravitational pulls of other planets across our solar system, the subtle yet significant non-zero quadrupole moment of the Sun, and the precession of the Earth's spring equinox, which serves as the reference axis. Despite these contributing factors, they collectively fall short of accounting for a deficit signified by
\begin{equation}\label{wobs}
    \left(\frac{dw}{dt} \right)_{\rm sec} = (42.9799 \pm 0.0009)''/\mbox{century}.
\end{equation}
The aforementioned figures were meticulously extracted from the latest estimations gleaned from data collected by the MESSENGER spacecraft \cite{Park_2017}.

For a numerical estimation of \eqref{wsec} we use $G = 6.674\times 10^{-11}$\,m$^3$/kg$\cdot$s$^2$, $M = M_{\odot} = 1.988\times 10^{30}\,$kg as the mass of the Sun, Mercury's eccentricity and semi-major axis $e = 0.205631$ and $a = 57.909\times10^9$\,m. Therefore, the equality between \eqref{wsec} and \eqref{wobs} implies that
\begin{equation}
    \kappa \approx 3.3\times 10^{-17} \, \mbox{(m/s)}^{-2}.
\end{equation}

The consequences of the constraints found in this section and the section before are discussed in the following.


\section{Final Remarks}

This study aims to extend conventional Newtonian gravitational theory by incorporating two scalar fields within a more generalised Lagrangian function and with well-behaved consequences to known results in celestial mechanics. The goal was to investigate how an extended Newtonian gravity would introduce effects that are usual in relativistic theories (under the domain of weak gravitational fields).

Specifically, we analysed the periastron advance of a binary system, and also the implications the theories bring for the equivalence principle. These effects give rise to observational constraints on the free functions of the theory, encoded in the parameter $\kappa$ [cf. equation \eqref{kappa}]. Data from the Nordtvedt effect suggests an upper limit for $\kappa$ of approximately $10^{-22}$ (m/s)$^{-2}$. On the other hand, for Mercury's perihelion shift the following indication emerges: $\kappa \cong - 3.3 \times 10^{-17}$ (m/s)$^{-2}$. Therefore, it becomes clear that it is impossible to formulate a coherent theory that simultaneously satisfies both of these critical constraints.

This conclusion should not necessarily be interpreted as a limitation of this class of models. The primary interest in exploring extensions of Newtonian gravity lies in their ability to serve as effective frameworks that are conceptually simpler than their relativistic counterparts and, at the same time, useful as a first approximation to modelling astrophysical systems. These extensions provide valuable insights when investigating various weak gravitational systems of particular interest, particularly in scenarios where relativistic effects can be neglected. It is important to recognise that a Newtonian theory is not expected to accurately describe the full spectrum of gravitational phenomena, given its simplifications. Thus, this work should be viewed as a fundamental study aimed at mapping how these extensions of Newtonian gravity can reproduce known effects observed in metric theories. Hence, the research proposed here seeks to bridge the gap between classical and relativistic treatments of gravity, elucidating the contexts in which simpler models remain relevant and useful.

The extension of Newtonian gravity considered in this work introduces corrections to the standard gravitational potential, $U$, that are proportional to $U^2$ and $\Phi_2$ [cf. equation \eqref{Ueff}]. Both corrections are characteristic features of metric theories of gravity and correspond to standard potentials within the parametrized post-Newtonian (PPN) formalism \cite{Will_2018}. It is worth noting that no Yukawa-like correction arises in this framework, which is a direct consequence of not including potential terms for any of the scalar fields in the Lagrangian. As a result, the class of theories examined here does not coincide with massive scalar–tensor theories in their weak-field limit. Since it is known that a Yukawa potential also leads to a periastron advance effect \cite{Karam:2026sqg}, this provides a strong motivation for incorporating such an analysis in future investigations.

Despite the generality of our analysis, it is worth mentioning two specific applications. The first one replicates the results of Ref.  \cite{Escorcio:2023}, by using the functions $k=1,\, f=\psi/\sigma^2, \,g=c^4$ and $h=\sigma$, which leads to $\kappa=\psi_0/c^4\omega$. Recently, Ref. \cite{Escorcio:2025} explored a new framework to describe cosmological expansion using a model of extended Newtonian gravity which is obtained by setting $k=1/\sigma, f=\psi/\sigma^2, g=c^4$ and $h=1$. As one can see, it is a simple modification of the Lagrangian worked out in Ref. \cite{Escorcio:2023} which clarifies the role of $\sigma$ in making the gravitational coupling $G$ variable. Notwithstanding, this model yields $\kappa=0$, since the $h$ function is constant, resulting in no post-Newtonian correction in the gravitational potential. This final case is particularly useful for demonstrating explicitly that, if the $\sigma$ field does not couple to matter, it exerts no influence on the motion of bodies up to first post-Newtonian order. Deviations from standard Newtonian theory would then arise only at the post-post-Newtonian order in the expansion. Although such higher-order perturbations are negligible for Solar System dynamics, they could become relevant in more precise or strongly relativistic systems, such as the double pulsar.

Two additional cases should also be cited, given their historical significance and due to their similarity to the efforts made by Einstein and Nordström to generalise Newton's theory within the framework of relativity \cite{Norton:1993}. By setting \( g = c^2 \), \( f = 1 \), and considering \( \psi \) as a unitary constant, we derive a field theory governed by the equation \( \Box\sigma = (-{4\pi G_0}/{\omega c^2}) h_\sigma \rho \). Aside from the constancy of the speed of light, Einstein's field equation is reproduced with \( h = \sigma \), while Nordström's formulation utilises \( h = \ln\sigma \) \cite{Einstein:1907,Nordstrom:1913}. The latter leads to $\kappa=-1$, while the former implies $\kappa=0$. Therefore, both formulations are inconsistent with Mercury's perihelion shift, although Einstein's would still preserve the equality between inertial and gravitational masses.

\acknowledgments{The authors thank FAPES, CAPES and CNPq for financial support. JDT specifically acknowledges FAPES for supporting this research through grant number 1016/2025 - P: 2025-QSS45.}

\section*{Data availability statement}
No data sets were generated during the current study. This is a theoretical work based on analytical and numerical calculations; all results are reproducible from the equations presented in the manuscript.

\bibliography{Refs}

\end{document}